\documentclass[prl,reprint,showpacs,amsmath,amssymb,superscriptaddress,longbibliography]{revtex4-1}
\usepackage{graphicx,bm,color,mathptmx}
\usepackage[colorlinks=true,linkcolor=blue,citecolor=blue,urlcolor =blue]{hyperref}

\newcommand{\Tr}{\mathop{\mathrm{Tr}} \nolimits}

\begin{document}

\title{Multiparameter Quantum Metrology of Incoherent Point Sources: Towards
  Realistic Superresolution} 

\author{J. \v{R}eha\v{c}ek} 
\affiliation{Department of Optics,
 Palack\'y University, 17. listopadu 12, 771 46 Olomouc, 
Czech Republic}

\author{Z. Hradil}
\affiliation{Department of Optics,
 Palack\'y University, 17. listopadu 12, 771 46 Olomouc, 
Czech Republic}

\author{B. Stoklasa} 
\affiliation{Department of Optics,
 Palack\'y University, 17. listopadu 12, 771 46 Olomouc, 
Czech Republic}

\author{M. Pa\'{u}r} 
\affiliation{Department of Optics,
 Palack\'y University, 17. listopadu 12, 771 46 Olomouc, 
Czech Republic}

\author{J. Grover} \affiliation{ESA---Advanced
  Concepts and Studies Office, European Space Research Technology
  Centre (ESTEC), Keplerlaan 1, Postbus 299, NL-2200AG Noordwijk,
  Netherlands}

\author{A. Krzic} 
\affiliation{ESA---Advanced  Concepts and Studies Office, 
 European Space Research Technology Centre (ESTEC), 
 Keplerlaan 1, Postbus 299, NL-2200AG Noordwijk, Netherlands}

\author{L. L. S\'{a}nchez-Soto} 
\affiliation{Departamento de \'Optica,
  Facultad de F\'{\i}sica, Universidad Complutense, 
28040~Madrid,  Spain} 
\affiliation{Max-Planck-Institut f\"ur die Physik des Lichts,
  G\"{u}nther-Scharowsky-Stra{\ss}e 1, Bau 24, 91058 Erlangen,
  Germany}

\begin{abstract}
  We establish the multiparameter quantum Cram\'er-Rao bound for
  simultaneously estimating the centroid, the separation, and the
  relative intensities of two incoherent optical point sources using
  a linear imaging system. For equally bright sources, the Cram\'er-Rao
  bound is independent of the source separation, which confirms that
  the Rayleigh resolution limit is just an artifact of the
  conventional direct imaging and can be overcome with an adequate
  strategy.  For the general case of unequally bright sources, the
  amount of information one can gain about the separation falls to
  zero, but we show that there is always a quadratic improvement in an
  optimal detection in comparison with the intensity
  measurements. This advantage can be of utmost important in realistic
  scenarios, such as observational astronomy.
\end{abstract}

\maketitle

The time-honored Rayleigh criterion~\cite{Rayleigh:1879ab} specifies
the minimum separation between two incoherent optical sources using a
linear imaging system. As a matter of fact, it is the size of the
point spread function~\cite{Goodman:2004aa} that determines the
resolution: two points closer than the PSF width will be difficult to
resolve due to the substantial overlap of their images.

Thus far, this Rayleigh criterion has been considered as a fundamental
limit.  Resolution can only be improved either by reducing the
wavelength or by building higher numerical-aperture optics, thereby
making the PSF narrower.  Nonetheless, outstanding methods have been
developed lately that can break the Rayleigh limit under special
circumstances~\cite{Dekker:1997aa,Betzig:2006aa,
  Hess:2006aa,Hell:2007aa,Kolobov:2007aa,Natsupres:2009aa,
  Hell:2009aa,Patterson:2010aa,Hemmer:2012aa,Cremer:2013aa}.  Though
promising, these techniques are involved and require careful control
of the source, which is not always possible, especially in
astronomical applications.

Despite being very intuitive, the common derivation of the Rayleigh
limit is heuristic and it is deeply rooted in classical optical
technology~\cite{Abbe:1873aa}.  Recently, inspired by ideas of quantum
information, Tsang and
coworkers~\cite{Tsang:2016aa, Nair:2016aa, Ang:2016aa,Tsang:2017aa}
have revisited this problem using the Fisher information and the
associated Cram\'er-Rao lower bound (CRLB) to quantify how well the
separation between two point sources can be estimated.  When only the
intensity at the image is measured (the basis of all the conventional
techniques), the Fisher information falls to zero as the separation
between the sources decreases and the CRLB diverges accordingly; this
is known as the Rayleigh curse~\cite{Tsang:2016aa}. However, when the
Fisher information of the complete field is calculated, it stays
constant and so does the CRLB, revealing that the Rayleigh limit is
not essential to the problem.

These remarkable predictions prompted a series of experimental
implementations~\cite{Paur:2016aa,Yang:2016aa, Tham:2016aa} and
further generalizations~\cite{Lupo:2016aa,Krovi:2016aa,
Rehacek:2017aa,Kerviche:2017aa,Zhuang:2017aa}, including the 
related question of source localization~\cite{Tsang:2015aa, 
Nair:2016ab,Sheng:2016aa}.  
All this previous work has focused on the estimation of the
separation, taking for granted a highly symmetric configuration with
identical sources.  In this Letter, we approach the issue in a more
realistic scenario, where both sources may have unequal
intensities. This involves the simultaneous estimation of separation,
centroid, and intensities.  Typically, when estimating multiple
parameters, there is a trade-off in how well different parameters may
be estimated: when the estimation protocol is optimized from the point
of view of one parameter, the precision with which the remaining ones
can be estimated deteriorates.

Here, we show that including intensity in the estimation problem does
lead to a reduction in the information for unbalanced sources. However
the information available in an optimal measurement still surpasses
that of a conventional direct imaging scheme by a significant margin
at small separations.  This suggests possible applications, for
example, in observational astronomy, where sources typically have small
angular separations and can have large differences in brightness.

Let us first set the stage for our simple model. We assume
quasimonochromatic paraxial waves with one specified polarization and
one spatial dimension, $x$ denoting the image-plane coordinate. The
corresponding object-plane coordinates can be obtained via the lateral
magnification of the system, which we take to be linear spatially
invariant~\cite{Goodman:2004aa}. 

To facilitate possible generalizations, we phrase what follows in a
quantum parlance. A wave of complex amplitude $U( x )$ can thus be
assigned to a ket $| U \rangle $, such that $U( x )= 
\langle x | U \rangle$, where $| {x} \rangle$ is a vector
describing a point-like source at ${x}$.

The system is characterized by its PSF, which represents its
normalized intensity response to a point source.  We denote this PSF
by $I (x) = | \langle x | \Psi \rangle |^{2} = |\Psi (x)|^{2}$, so that
$\Psi(x)$ can be interpreted as the amplitude PSF.

Two incoherent point sources, of different intensities and separated
by a distance $\mathfrak{s}$, are imaged by that system. The signal
can be represented as a density operator
\begin{equation}
  \varrho_{\bm{\theta}}= \mathfrak{q} \, \varrho_{+} 
  + (1-\mathfrak{q} ) \, \varrho_{-} \, ,
\end{equation}
where $\mathfrak{q}$ and $1 - \mathfrak{q}$ are the intensities of the
sources, with the proviso that  the total intensity is normalized to
unity.  In addition, we have defined $\varrho_{\pm}=
|\Psi_{\pm}\rangle \langle\Psi_{\pm}|$ and the $x$-displaced PSF states are
\begin{equation}
  \langle x | \Psi_{\pm} \rangle = 
  \langle x - \mathfrak{s}_{0} \mp \mathfrak{s}/2 | \Psi  \rangle = 
  \Psi(x-\mathfrak{s}_{0} \mp \mathfrak{s}/2),
\end{equation}
so that they are symmetrically located around the geometric centroid
$\mathfrak{s}_{0} = \tfrac{1}{2} (x_{+} + x_{-})$. Note that
\begin{equation}
  | \Psi_{\pm} \rangle = \exp[ - i (\mathfrak{s}_{0} \pm \mathfrak{s}/2) P ] 
  | \Psi \rangle \, ,
\end{equation}
where $P$ is the momentum operator, which generates displacements in
the $x$ variable. As in quantum mechanics, it acts as a derivative
$P = - i \partial_{x}$.  These spatial modes are not orthogonal
($\langle \Psi_{-} | \Psi_{+} \rangle \neq 0$), so they cannot be
separated by independent measurements.

The density matrix $ \varrho_{\bm{\theta}}$ gives the normalized mean
intensity: $ \varrho_{\bm{\theta}} (x) = \mathfrak{q} \, {|\Psi(x-
  \mathfrak{s}_{0} - \mathfrak{s}/2)|^2} + (1-\mathfrak{q})\, |\Psi(x
-\mathfrak{s}_{0} + \mathfrak{s}/2)|^2$, and  depends on the centroid 
$\mathfrak{s}_{0}$, the separation $\mathfrak{s}$, and the relative
intensities of the sources $\mathfrak{q}$. This is indicated by the
vector $\bm{\theta} = (\mathfrak{s}_{0}, \mathfrak{s}, \mathfrak{q})^{t}$.
The task is to estimate the values of $\bm{\theta}$ through the
measurement of some observables on $\varrho_{\bm{\theta}}$. In turn, a
quantum estimator $\widehat{\bm{\theta}}$ for $\bm{\theta}$ is a
selfadjoint operator representing a proper measurement followed by
data processing performed on the outcomes. Such a parameter estimation
implies an additional uncertainty for the measured value, which
cannot be avoided. 

In this multiparameter estimation scenario, the central quantity is
the quantum Fisher information matrix (QFIM)~\cite{Petz:2011aa}. This
is a natural generalization of the classical Fisher information, which is a
mathematical measure of the sensitivity of an observable quantity (the
PSF, in our case) to changes in its underlying parameters. However,
the QFIM it is optimized over all the possible quantum
measurements. It is define
reads   
\begin{equation}
Q_{\alpha \beta} (\bm{\theta}) = \tfrac{1}{2} \Tr (
\varrho_{\bm{\theta}} \{ L_{\alpha}, L_{\beta} \}) \, ,
\end{equation} 
where the Greek indices run over the components of the vector
$\bm{\theta}$ and $\{ \cdot, \cdot \}$ denotes the anticommutator. 
Here, $L_{\alpha}$ stands for the symmetric logarithmic
derivative~\cite{Helstrom:1967aa} with respect  the parameter
$\theta_{\alpha}$, defined implicitly by  
$\tfrac{1}{2} (L_{\alpha} \varrho_{\bm{\theta}} +
\varrho_{\bm{\theta}} L_{\alpha} ) = \partial_{\alpha}
\varrho_{\bm{\theta}}$, with $\partial_{\alpha} = \partial/ \partial
\theta_{\alpha}$.   

Upon writing $\varrho_{\bm{\theta}}$ in its eigenbasis
$\varrho_{\bm{\theta}} = \sum_{n} \lambda_{n} | \lambda_{n}\rangle
\langle \lambda_{n}| $, the QFIM per detection event can be concisely
expressed as~\cite{Paris:2009aa}
\begin{equation}
  Q_{\alpha \beta} (\bm{\theta}) = 
  2 \sum_{m,n}\frac{1}{\lambda_m+\lambda_n}
  \langle \lambda_m|  \partial_{\alpha} \varrho_{\bm{\theta}}
  |\lambda_n\rangle  \langle \lambda_n|
  \partial_{\beta} \varrho_{\bm{\theta}} |\lambda_m\rangle \, ,
\end{equation}
and the summation extends over $m,n$ with $\lambda_m+\lambda_n\neq 0$.
In addition, the constraints of unity trace $\sum_m \lambda_m=1$ and
the completeness relation
$\sum_m |\lambda_m\rangle\langle \lambda_m|=\openone$ have to be
imposed.

 The QFIM is a distinguishability metric on the space of quantum
 states and leads to the multiparameter quantum
 CRLB~\cite{Braunstein:1994aa,Szczykulska:2016aa}: 
\begin{equation}
\mathrm{Cov} (\widehat{\bm{\theta}} ) \ge
Q^{-1}  (\bm{\theta} )  \, ,
\end{equation}
where $\mathrm{Cov} (\widehat{\bm{\theta}} ) = \mathbb{E} [
(\widehat{\theta}_{\alpha} - \theta_{\alpha})
(\widehat{\theta}_{\beta} - \theta_{\beta}) ] $ refers to the
covariance matrix for a locally unbiased estimator
$\widehat{\bm{\theta}}$ of the quantity $\bm{\theta}$ and 
$\mathbb{E} [Y]$ is the expectation value of the random variable $Y$. 
In particular,  the individual parameter $\theta_{\alpha}$ can be
estimated with a variance satisfying $\mathrm{Var}
(\widehat{\theta}_{\alpha}) \ge  (Q^{-1})_{\alpha \alpha}  (\bm{\theta}
)$, and a positive operator-valued  measurement (POVM)  attaining this
accuracy is given by the eigenvectors of $L_{\alpha}$.  Unlike for
a single parameter,  the collective bound is not always
saturable: the intuitive reason for this is incompatibility of the
optimal measurements for different
parameters~\cite{Holevo:2003fv}.

If the  operators $L_{\alpha}$ corresponding to the different
parameters commute, there is no additional difficulty in extracting
maximal information from a state on all parameters simultaneously. If
they do not commute, however, this does not immediately imply that it
is impossible to simultaneously extract information on all parameters
with precision matching that of the separate scenario for each.
As discussed in a number of
papers~\cite{Matsumoto:2002aa,Crowley:2014aa,Ragy:2016aa}  the
multiparameter quantum CRLB can be saturated provided 
\begin{equation}
\label{eq:comp}
\Tr ( \varrho_{\bm{\theta}}  [L_{\alpha} ,L_{\beta} ]) = 0 \, .
\end{equation}   
Then, optimal measurements can be found by optimizing over the
classical Fisher information, as the QFIM is an upper bound for the
former quantity.  This can be efficiently accomplished by global
optimization algorithms~\cite{Global:1995aa}. For our particular case,
it is easy to see that the condition \eqref{eq:comp} is fulfilled
whenever the PSF is real, $\Psi(x)^\ast=\Psi(x)$, which will be
assumed henceforth.

To proceed further, we note that the density matrix
$\varrho_{\bm{\theta}}$ is, by definition, of rank 2. The QFIM reduces
then to the simpler form
\begin{eqnarray}
  \label{twocomponents}
 Q_{\alpha\beta}  = &  - & \frac{3}{\lambda_{1}}  
\langle \lambda_{1}| 
\partial_{{\alpha}} \varrho_{\bm{\theta}} 
| \lambda_{1} \rangle \langle \lambda_{1}| 
\partial_{\beta} \varrho_{\bm{\theta}}  |\lambda_{1} \rangle 
\nonumber  \\
& - & \frac{3}{\lambda_{2}}  
\langle \lambda_{2}| 
\partial_{{\alpha}} \varrho_{\bm{\theta}} 
| \lambda_{2} \rangle \langle \lambda_{2}| 
\partial_{\beta} \varrho_{\bm{\theta}}  |\lambda_{2} \rangle 
\nonumber \\
& +& 4 \left(1-\frac{1}{\lambda_1} -\frac{1}{\lambda_2} \right )
\langle \lambda_1 |
\partial_{{\alpha}} \varrho_{\bm{\theta}}  |
\lambda_2\rangle  \langle \lambda_2|
\partial_{{\beta}} \varrho_{\bm{\theta}}  |\lambda_1\rangle
\nonumber \\
    &+& \frac{4}{\lambda_1} \langle \lambda_1| \partial_{\alpha}
      \varrho_{\bm{\theta}} \partial_{\beta} \varrho_{\bm{\theta}}
        |\lambda_1 \rangle +
  \frac{4}{\lambda_2}\langle  \lambda_2|\partial_{\alpha}
      \varrho_{\bm{\theta}} \partial_{\beta} \varrho_{\bm{\theta}}
        |\lambda_2\rangle \, . \quad
\end{eqnarray}
The derivatives involved in this equation can be easily evaluated;
the result reads
\begin{equation}
  \label{derivatives}
  \begin{split}
\partial_{\mathfrak{s}_{0}} \varrho_{\bm{\theta}} &=
 i[ \varrho_{\bm{\theta}} , P] \, , \\
 \partial_{\mathfrak{s}}  \varrho_{\bm{\theta}} & = 
\tfrac{i}{2} ( \mathfrak{q} [  \varrho_{+}, P] -
  (1- \mathfrak{q} )[\varrho_{-}, P] ) \, , \\
    \partial_{\mathfrak{q}}  \varrho_{\bm{\theta}} &=\varrho_{+} -
    \varrho_{-} \, .
  \end{split}
\end{equation}
To complete the calculation it proves convenient to write the two
nontrivial eigenstates of $\varrho_{\bm{\theta}}$ in terms of
non-orthogonal component states $|\Psi_{\pm}\rangle$:
$|\lambda_{1,2}\rangle = a_{1,2} |\Psi_{+} \rangle + b_{1,2}
|\Psi_{-}\rangle$, where $a_{1,2} $ and $b_{1,2} $ are easy-to-find
yet complex functions of the separation and the intensities and whose
explicit form is of no relevance for our purposes here.  Substituting
this and Eq.~\eqref{derivatives} into Eq.~\eqref{twocomponents}, and
after a lengthy calculation, we obtain a compact expression for the
QFIM
\begin{equation}
  \label{qfi}
  Q = 4 
  \left(
    \begin{array}{ccc}
      p^{2} + 4 \mathfrak{q} (1- \mathfrak{q}) \wp^{2} 
     & (\mathfrak{q}-1/2) p^{2}
      & -i w  \wp \\
      ( \mathfrak{q}-1/2) p^{2} & p^{2}/4 & 0\\
      -i w \wp & 0 & \displaystyle  \frac{1-w^2}
   {4 \mathfrak{q}(1- \mathfrak{q})}
    \end{array}
  \right) \, .
\end{equation}
This is our central result. The QFIM depends only on the following quantities
\begin{eqnarray} 
w & \equiv & \langle \Psi_{\pm} |\Psi_{\mp} \rangle = 
\langle \Psi |\exp(i \mathfrak{s} P) |\Psi \rangle \, , 
\nonumber  \\ 
p^{2} & \equiv & \langle \Psi_{\pm} |P^2 |\Psi_{\pm}\rangle 
= \langle \Psi | P^2 | \Psi\rangle \, ,  \\
\wp & \equiv& \pm \langle \Psi_{\pm} | P| \Psi_{\mp}\rangle =
\langle \Psi|\exp(i \mathfrak{s} P) P|\Psi\rangle \, . \nonumber 
\end{eqnarray}
Interestingly, $p^{2}$ is fully determined by the shape of the PSF,
whereas both $w$ and $\wp$  depend on the separation
$\mathfrak{s}$. Furthermore, $\wp$ is purely imaginary.

In what follows, rather than the variances themselves, we will use the
inverses ${H}_{\alpha} = 1/ \mathrm{Var} (\theta_\alpha)$,
usually called the precisions~\cite{Bernardo:2000aa}. In this way, we
avoid potential divergences at $\mathfrak{s}=0$.

\begin{figure}
  \centerline{
    \includegraphics[width=0.85\columnwidth]{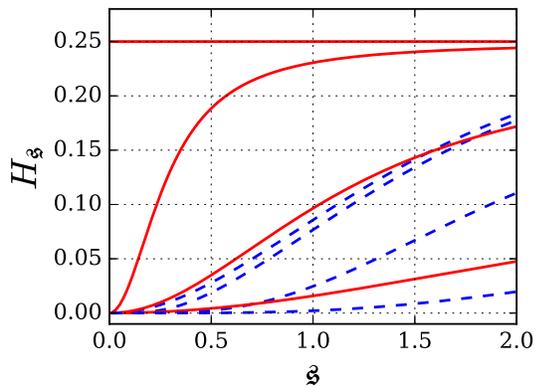}}
  \caption{Precision $H_{\mathfrak{s}}$ in the separation
    $\mathfrak{s}$ as inferred by optimal (red solid lines) and direct
    (blue broken lines) detections for different relative intensities
    of the two sources. The values of $\mathfrak{q}$, from top to
    bottom, are 0.5, 0.45, 0.3, and 0.1. Notice that the performance
    of the optimal detection is rather sensitive to small deviations
    from equal brightness over a wide range of separations.}
  \label{fig1}
\end{figure}

The QFIM~\eqref{qfi} nicely shows the interplay between various signal
parameters. First, notice that $Q$ is independent of the centroid, as
might be expected. Second, for equally bright sources,
$\mathfrak{q}=1/2$, the measurement of separation $\mathfrak{s}$ is
uncorrelated with the measurements of the remaining parameters and we
have $H_{\mathfrak{s}}( \mathfrak{q}=1/2)=p^{2}$, a well known result,
and the Rayleigh curse is lifted~\cite{Paur:2016aa}.  This happy
coincidence does not hold for unequal intensities
$\mathfrak{q} \neq 1/2$; now, the separation is correlated with the
centroid (via the intensity term $\mathfrak{q}-1/2$) and the centroid
is correlated with the intensity (via $p^{2}$). This can be
intuitively understood: unequal intensities result in asymmetrical
images and finding the centroid is no longer a trivial task. This
asymmetry, in turn, depends on the relative brightness of the two
components. Hence, all the three parameters become entangled and, as
we shall see, having separation-independent information about
$\mathfrak{s}$ is no longer possible.

By inverting the QFIM we immediately get
\begin{equation}
  H_{\mathfrak{s}} = p^{2} 
  \frac{\mathcal{Q}^{2} \wp^{2} + \mathcal{Q}^{2} p^{2}(1-w^2)}
  {\mathcal{Q}^{2}   \wp^{2}+ p^{2}(1-w^2)} \, , 
\end{equation}
where $ 0\le \mathcal{Q}^{2}  \equiv 4 \mathfrak{q}(1-\mathfrak{q}) \le 1$.
Obviously, $H_{\mathfrak{s}} (\mathfrak{q} ) \le H_{\mathfrak{s}}
(\mathfrak{q} =1/2) =p^{2} $ and $\lim\limits_{\mathfrak{q} \rightarrow 0,1}
H_{\mathfrak{s}}(\mathfrak{q} )=0$, which demonstrates that resolving 
two highly unequal sources is difficult, even at the quantum limit.

The instance of large brightness differences and  small separations is
probably the most interesting regime encountered, e. g., in exoplanet 
observations. We first expand the $\mathfrak{s}$-dependent quantities:
\begin{equation}
  \begin{split}
    w( \mathfrak{s} )& =\langle \Psi|e^{i \mathfrak{s} P}| \Psi \rangle
    \simeq 1 - \tfrac{1}{2} p^{2} \mathfrak{s}^2 + \tfrac{1}{24} p^{4}
    \mathfrak{s}^4 ,\\
    p (\mathfrak{s} )&= \langle \Psi|P e^{i \mathfrak{s} P}|\Psi\rangle 
  \simeq  i p^{2} \mathfrak{s} - i \tfrac{1}{6} p^{4} 
  \mathfrak{s}^{3} \, ,
  \end{split}
\end{equation}
where $p^{4} =\langle \Psi|P^4| \Psi\rangle$ is the fourth moment of the
PSF momentum. Then, as $\mathfrak{s}\ll 1$, we get (for $0 <
\mathcal{Q} < 1$) 
\begin{equation}
  \label{approx}
  \begin{split}
    H_{\mathfrak{s}_{0}} & \simeq  \mathcal{Q}^{2}\,
   \mathrm{Var} (\hat{P}^{2}) \,  \mathfrak{s}^{2} \, ,\\
    H _{\mathfrak{s}}& \simeq  \frac{\mathcal{Q}^{2}}
    {4(1-\mathcal{Q}^{2})} \, \mathrm{Var} (\hat{P}^{2})
    \, \mathfrak{s}^2, \\
    H_{\mathfrak{q}} & \simeq \frac{1}{\mathcal{Q}^{2}} 
   \mathrm{Var} (\hat{P}^{2}) \,  \mathfrak{s}^4\, .
  \end{split}
\end{equation}
The  PSF enters these expressions  through the variance of
$P^2$:  $ \mathrm{Var}(\hat{P}^2)= p^{4}-
p^{2}$. This leaves room for optimization, provided the PSF can be
controlled. For a fixed PSF, the information about all three
parameters apparently vanishes with $\mathfrak{s} \rightarrow 0$
unless $\mathfrak{q} =1/2$. And since exactly balanced sources never
happen, the information about very small separations always drops to
near zero and the Rayleigh curse is unavoidable. However, significant
improvements of the optimal measurement schemes over the standard
intensity detection are still possible.

\begin{figure}
  \centerline{
    \includegraphics[width=0.90\columnwidth]{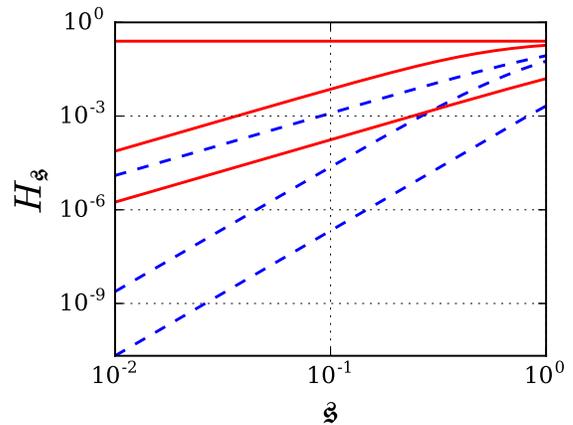}}
  \caption{Precisions as in Fig.~\ref{fig1}, but visualized on a
    logarithmic scale. Slopes of straight lines translate to the
    powers of $\mathfrak{s}$. The values of $\mathfrak{q}$ are, from
    top to bottom, 0.5, 0.4, and 0.1.}
 \label{fig2}
\end{figure}

To illustrate this point let us consider a Gaussian response
$\langle x|\Psi\rangle=(2\pi)^{(1/4)}\exp ( -x^2/4)$ of unit width,
which will serve from now on as our basic unit length. We shall
compare the quantum limit given by \eqref{qfi} with that given by the
classical Fisher information for the direct intensity measurement.  We
assume no prior knowledge about any of the three parameters.

Figure~\ref{fig1} plots information about separation
$H_{\mathfrak{s}}$ for different relative intensities
$\mathfrak{q}$. Unbalanced intensities make both optimal and intensity
detection go to zero for small separations, however the former at a
much slower rate. Hence, optimal information to intensity information
increases  with decreasing separations, regardless of whether the
sources are balanced. 

The reason becomes obvious with the same data visualized on the
logarithmic scales, as shown in Fig.~\ref{fig2}. In the region of
$\mathfrak{s}\ll \sigma$, we can discern two regimes of
importance. For balanced sources,
$H_{\mathfrak{s}}^{\textrm{opt}} \propto 1$ and
$H_{\mathfrak{s}}^\textrm{int}\propto \mathfrak{s}^2$. For unbalanced
sources, $H_{\mathfrak{s}}^{\textrm{opt}} \propto \mathfrak{s}^2$, as
we have seen, and
$H_{\mathfrak{s}}^\textrm{int}\propto \mathfrak{s}^4$. In consequence,
there is always a factor of $\mathfrak{s}^{-2}$ improvement of the
optimal detection over the standard one, irrespective of the true
values of the signal parameters. In practice, this means that when we
already are well below the Rayleigh limit, if we decrease the
separation $10$ times, about $10,000$ times more photons must be
detected with a CCD camera to keep the accuracy of the measurement,
while only $100$ times more would suffice for optimal
measurement. This amounts to saving $99\%$ of detection time with the
optimal detection scheme.

\begin{figure}
  \centerline{\includegraphics[width=0.85\columnwidth]{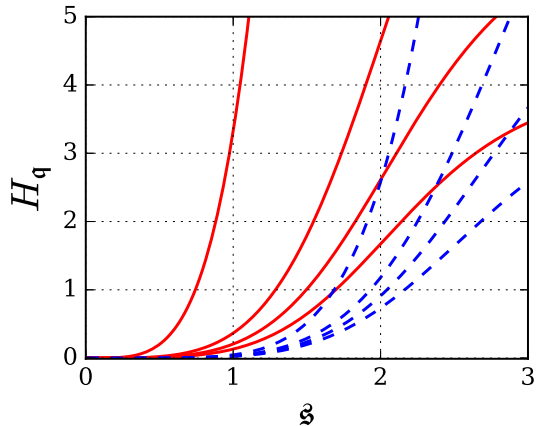}}
  \caption{Precision about relative intensity $\mathfrak{q}$ as
    inferred by the optimal detection (red solid lines) and the direct
    detection (blue broken lines) for different relative intensities
    of the two sources. The values of $\mathfrak{q}$, from
    bottom to top are  0.5, 0.2, 0.1, and 0.01.}
  \label{fig3}
\end{figure}

Finally, Fig.~\ref{fig3} presents a similar comparison now concerning
the information about the relative intensity $H_{\mathfrak{q}}$. Here,
optimal information and intensity information always scale as
$\mathfrak{s}^4$ and $\mathfrak{s}^6$, respectively, and the same
$\mathfrak{s}^{-2}$ gain in performance appears. Notice the reversed
ordering of curves with $\mathfrak{q}$, meaning that now, the
information increases rather than decreases with increasing intensity
difference, which reveals the complementarity between these
magnitudes. Also notice that the broken lines converge as we approach 
$\mathfrak{s}=0$. It can be shown that the leading term for intensity
detection for small separations is $p$-independent in contrast to the
optimal detection, which displays a strong
$H_{\mathfrak{q}}^\textrm{opt} \propto \mathfrak{q}^{-1}$ dependence
for $\mathfrak{q} \ll 1/2$. This highlights the advantage of an
optimal detection scheme for astronomical observations. For example
more than a quarter of catalogued binary systems consist of stars
that differ in brightness by more than an order of
magnitude~\cite{X}, and the darkest known exoplanet is three orders of
magnitude dimmer than its host star in the
infrared~\cite{Barclay:2012aa}.

In summary, we have presented a comprehensive analysis of the ultimate
precision bounds for estimating the centroid, the separation, and the
relative intensities of two pointlike incoherent sources. For equally
bright sources, the quantum Fisher information remains constant, which
translates into the fact that the Rayleigh limit is not essential and
can be lifted. On the other hand, for unequally bright sources, the
information about very small separations always drops to 
near zero and the Rayleigh curse is unavoidable. Nonetheless,
significant improvements can still be expected with optimal detection
schemes. 

We acknowledge financial support from the Technology Agency of the
Czech Republic (Grant TE01020229), the Grant Agency of the Czech
Republic (Grant No. 15-03194S), the IGA Project of the Palack{\'y}
University (Grant No. IGA PrF 2016-005), the European Space Agency's
ARIADNA scheme, and the Spanish MINECO (Grant FIS2015-67963-P).


%

\end{document}